\documentclass[twocolumn,aip,rsi,reprint]{revtex4-1}
\usepackage[latin9]{inputenc}
\usepackage{color}
\usepackage{amsmath}
\usepackage{amssymb}
\usepackage{graphicx}
\usepackage[unicode=true,pdfusetitle,
 bookmarks=false,
 breaklinks=false,pdfborder={0 0 0},pdfborderstyle={},backref=false,colorlinks=true]
 {hyperref}
\hypersetup{
 colorlinks,linkcolor=red,citecolor=blue}
\usepackage{breakurl}

\makeatletter

\newcommand*\LyXThinSpace{\,\hspace{0pt}}

\usepackage{amsfonts}

\makeatother

\begin{document}

\title{Polarization mode hybridization and conversion in phononic wire waveguides}

\author{Zhen Shen}
\thanks{These authors contributed equally to this work.}
\affiliation{Department of Electrical Engineering, Yale University, New Haven, CT 06520, USA}

\author{Wei Fu}
\thanks{These authors contributed equally to this work.}
\affiliation{Department of Electrical Engineering, Yale University, New Haven, CT 06520, USA}

\author{Risheng Cheng}
\affiliation{Department of Electrical Engineering, Yale University, New Haven, CT 06520, USA}

\author{Hendrick Townley}
\affiliation{Department of Electrical Engineering, Yale University, New Haven, CT 06520, USA}

\author{Chang-Ling Zou}
\affiliation{Department of Electrical Engineering, Yale University, New Haven, CT 06520, USA}

\author{Hong X. Tang}
\affiliation{Department of Electrical Engineering, Yale University, New Haven, CT 06520, USA}
\email{hong.tang@yale.edu}

\begin{abstract}
Phononic wire waveguides of subwavelength cross-section support two orthogonal polarization modes: the out-of-plane motion dominated Rayleigh-like and the in-plane motion dominated Love-like modes, analogous to transverse-electric and transverse-magnetic modes in photonic waveguides. Due to the anisotropic elasticity of the substrate material, the polarization states of phonons propagating along certain crystallographic orientations can strongly hybridize. Here we experimentally investigate the orientation-dependent mode hybridization in phononic wire waveguides patterned from GaN-on-sapphire thin films. Such mode hybridization allows efficient actuation of piezoelectrically inactive Love-like modes using common interdigital electrodes designed for Rayleigh-like modes, and further enables on-chip polarization conversion between guided transverse modes. Both are important for on-chip implementation of complex phononic circuits. 
\end{abstract}
\maketitle

Over the past few decades, a variety of surface acoustic wave (SAW) devices have been developed for radio frequency electronics and communication systems, including delay lines, filters, resonators and precision oscillators \citep{Book-1,Book-2}. SAW devices are also moving into
the lab-on-a-chip arena and find use in sensing applications \citep{Book-1,Book-2,Sensor2008,Sensor}. Lately, there have been emerging interests in the realization of light-sound
and spin-sound interactions using SAW \citep{M-Li,M-Li-optica,brillouin,WeiFuOE,Nat-Photonics,H-Wang}. However, in almost all of those applications, the SAW propagates on the chip surface in a two-dimensional (2D) or quasi-2D slab of many wavelength wide without tight lateral confinement. It is anticipated that the strength of these interactions can be greatly enhanced by using phononic strip waveguides of subwavelength cross-section or phononic wires, to confine phonon fields\citep{Wei,Wance}. Such high-confinement phononic circuits could perform functions that complement electronic and photonic circuits, at operation frequencies matching that of microwave and wavelengths matching that of optical
signals \citep{waveguide1,waveguide2,waveguide3,waveguide4,waveguide5}.

Similar to optical waveguides which support orthogonal polarization transverse-electrical (TE) and transverse-magnetic (TM) modes, phononic waveguides support two types of phonon modes: the vertically polarized (with dominant out-of-plane motion) Rayleigh-like mode and the horizontally polarized (with dominant in-plane motion) Love-like mode. While the Rayleigh-like mode is piezoelectrically active and has been extensively studied in literature and more commonly utilized in today's RF signal processing devices \citep{Book-2,Rayleigh}, the Love-like mode is particularly interesting for building high quality factor ($Q$) acoustic resonators and for sensing applications due to their relative immunity to air and fluid damping \citep{Sensor,resonator1,resonator2,SensorLove}.
However, it is difficult to actuate the Love-like mode, which is normally piezoelectrically inactive, by an interdigital transducer (IDT).

Here we study the phononic modes propagating along a high-confinement waveguide structure\textemdash GaN-on-sapphire phononic wire waveguide. Although GaN presents isotropic elastic properties in the \emph{c}-plane, the sixfold-symmetric elastic properties of \emph{c}-sapphire as the substrate gives rise to the anisotropy of SAW. With IDT optimized for Rayleigh-like mode excitation, we observe only Rayleigh-like mode for waveguides along $\left[1\bar{1}00\right]$ direction while both Rayleigh-like and Love-like modes are actuated in waveguides oriented along $\left[11\bar{2}0\right]$ direction. To experimentally verify the excitation of Love-like mode, a home-built phase sensitive vibrometer is utilized to image both the amplitude and phase pattern of the modes' out-of-plane displacement. The results unambiguously demonstrate the
hybridization between the two transverse modes, which is attributed to the anisotropic sapphire substrate and confirmed by numerical simulations. Such mode hybridization allows the actuation of the Love-like mode, not-directly addressable by the IDT, and achieves mode conversion between
the two guided transverse modes, which is of great interests for fundamental studies and future applications based on the phononic circuits.

\begin{figure}
\includegraphics[width=1\columnwidth]{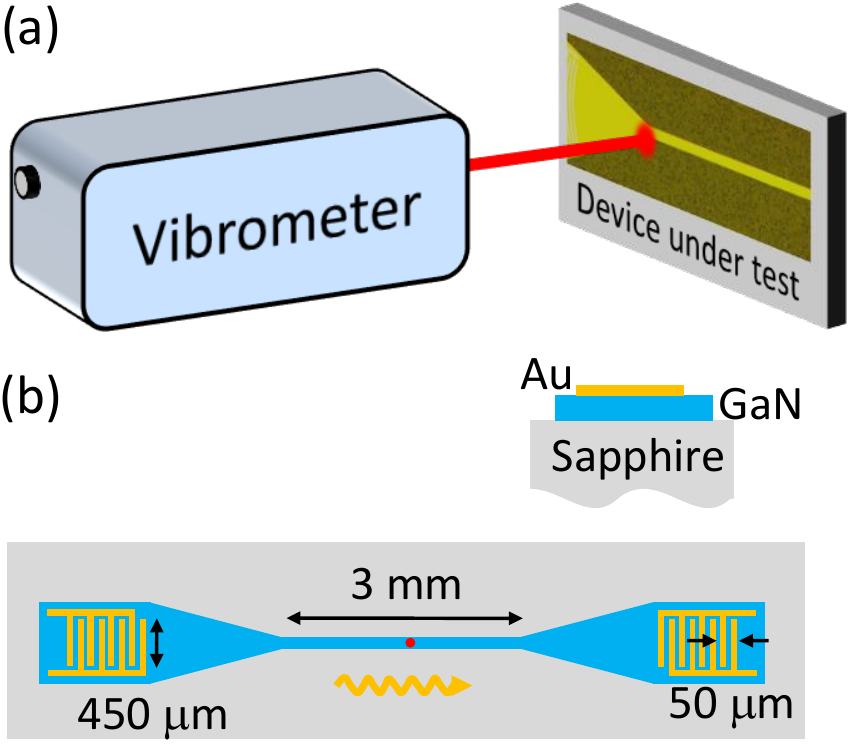}\caption{(a) A block diagram of our experimental setup together with a device under test. (b) The schematic of a typical phononic wire waveguide device. (b) is not to scale.}

\label{fig1}
\end{figure}

A block diagram of our experimental setup together with a device under test is shown in Fig.$\,$\ref{fig1}a. A home-built vibrometer is exploited to map the amplitude and phase of out-of-plane acoustic vibration. A similar setup has been used to realize the phase sensitive imaging of acoustic modes in high quality factor Fabry-Perot phonon resonators and 10 GHz thickness
vibrations in an AlN microdisk resonator and was described in detail in our previous work \citep{FP-SAW,10GHzvibrometer}. Briefly, the principle of the vibrometer is based on a quadrature measurement of vibration-modulated light signals by a heterodyne interferometer.
For vibration frequency of 115 MHz, the sensitivity of the vibrometer is around $70\:\textrm{fm}/\sqrt{\textrm{Hz}}$, which is mainly limited by electrical noise of photo-detector in the vibrometer.

\begin{figure}
\includegraphics[width=1\columnwidth]{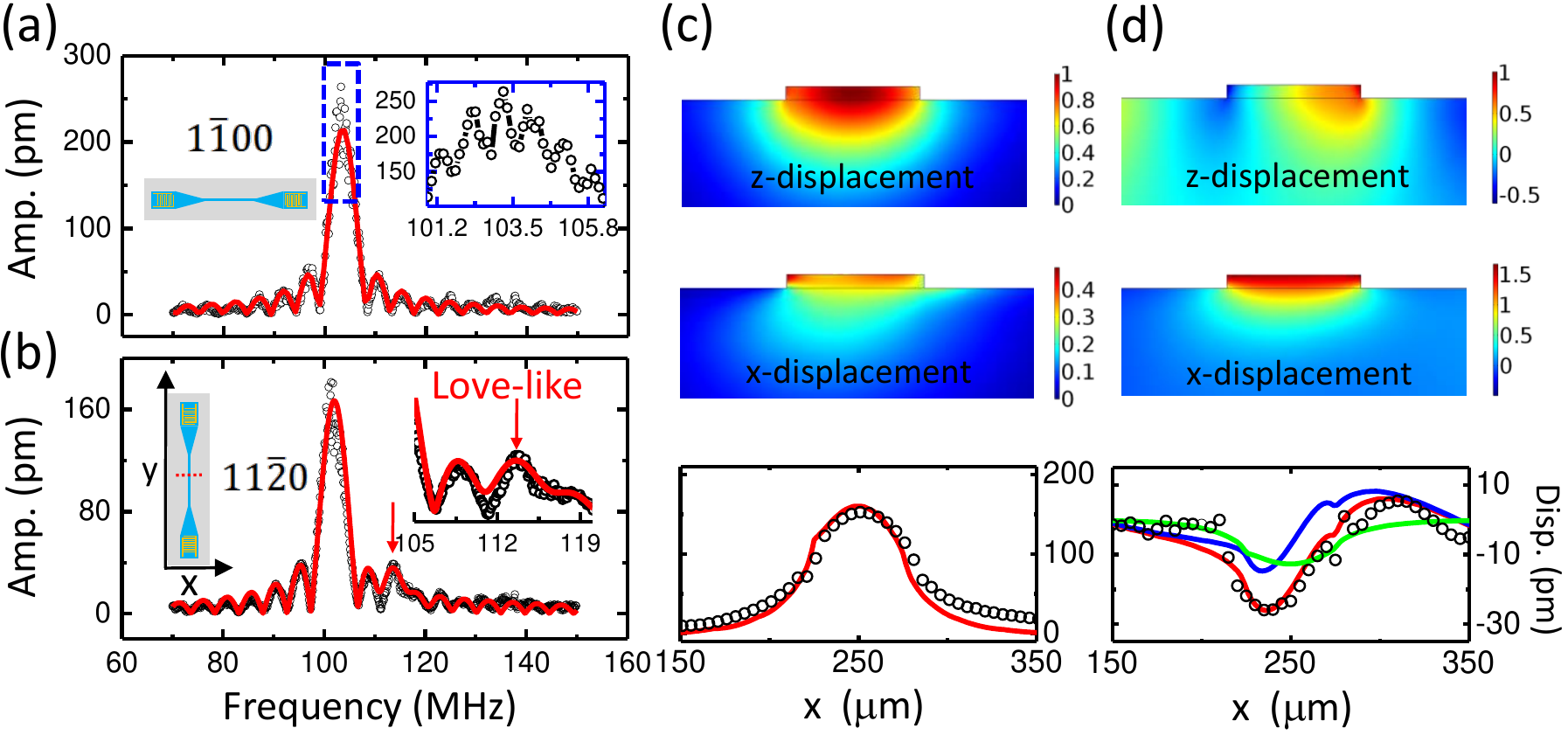}

\caption{(a-b) The frequency responses of IDT in the $\left[1\bar{1}00\right]$
and $\left[11\bar{2}0\right]$ direction, respectively. The red lines are sinc-function fit at a center frequency of 103.5 MHz in (a), 102 MHz and 113.5 MHz in (b). The configurations of the two orthogonal waveguides are illustrated in the insets of (a-b). Red dots in Fig.
\ref{fig1}(b) indicate the measurement point. The details of central peak associated with Rayleigh-like mode and Love-like mode are shown in the insets of (a-b), respectively. (c-d) Simulated x-displacement (in-plane) and z-displacement (out-of-plane) fields of the Rayleigh-like and Love-like mode propagating along the $\left[11\bar{2}0\right]$ direction, respectively. The bottom panels are the measured out-of-plane
displacement distributions across the waveguide taken at red line in the inset of (b). Red line in (c) represents the simulated result of Rayleigh-like mode. The blue line and green line in (d) represent the profile of pure Love-like mode and Rayleigh-like mode, respectively. The red line is the superposition of these two modes.}

\label{fig2}
\end{figure}

Figures$\,$\ref{fig1}(a-b) show the optical micrograph and schematic of a typical phononic waveguide device, which features IDTs patterned from 100-nm-thick gold and a 5-$\textrm{\ensuremath{\mu}m}$-thick wire waveguide patterned from GaN. The IDTs have an aperture of $450\:\textrm{\ensuremath{\mu}m}$
and a period of $50\:\textrm{\ensuremath{\mu}m}$, corresponding to the wavelength of the Rayleigh-like mode they excite. Two orthogonally
oriented $50\:\textrm{\ensuremath{\mu}m}$-wide waveguides are fabricated along the $\left[1\bar{1}00\right]$ and $\left[11\bar{2}0\right]$ in-plane direction of \emph{c}-sapphire substrate, as shown in the insets of Figs.$\,$\ref{fig2}(a-b).

Figures$\,$\ref{fig2}(a) and (b) show the frequency response of the two orthogonal waveguides with a fixed RF input power of around 27 dBm. The frequency responses are obtained while the signal laser spot is focused on a fixed spatial point on the waveguide and the excitation frequency is scanned from $70\:\textrm{MHz}$ to $150\:\textrm{MHz}$. Two transmission peaks appear in the spectra of the sample along the $\left[11\bar{2}0\right]$ direction. The lower-frequency peak is associated with the Rayleigh-like mode, while the other corresponds to the Love-like mode. In contrast, only the Rayleigh-like mode is observed for the sample along the $\left[1\bar{1}00\right]$ direction, where the frequency response displays a well-defined sinc-function curve. The peak frequency of the Rayleigh-like mode in the $\left[1\bar{1}00\right]$ direction is about 1.5 MHz higher than that in the $\left[11\bar{2}0\right]$ direction. This is due to the anisotropic substrate, which will be further discussed below. The small oscillations in the spectrum (right inset of Fig.$\,$\ref{fig2}(a)) are due to the interference between the forward phonon and the backward phonon bouncing between the IDTs.

\begin{figure}
\begin{centering}
\includegraphics[width=1\columnwidth]{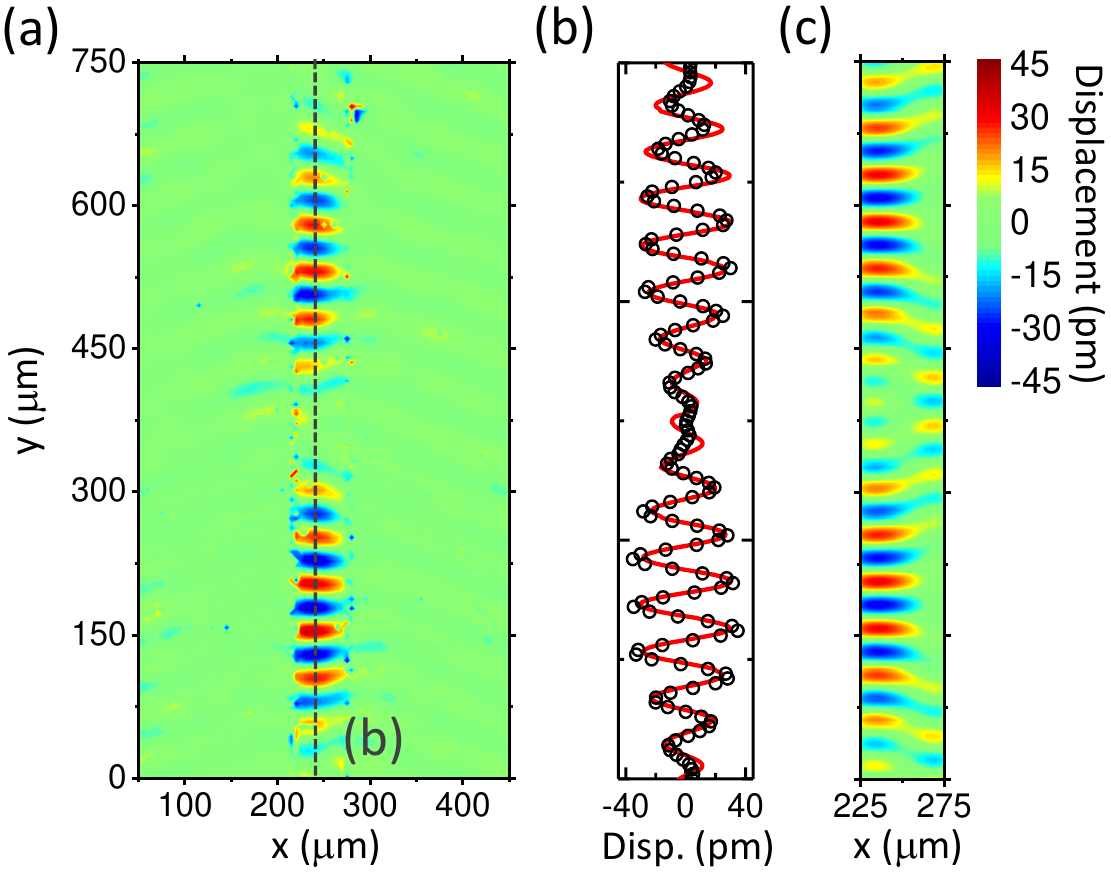}
\par\end{centering}
\caption{(a) Measured displacement field acquired at the excitation frequency of the Love-like mode, 113.5 MHz, with a detection bandwidth of 10 Hz. (b) Displacement distribution along the propagating direction (y-axis) approaching the center of the waveguide ($x=240\:\textrm{\ensuremath{\mu}m})$.
(c) Simulated displacement of a superposition of Love-like mode and Rayleigh-like mode, with the x ranging from $225\:\textrm{\ensuremath{\mu}m}$
to $275\:\textrm{\ensuremath{\mu}m}$. Red line in (b) is the simulation displacement.}

\label{fig3}
\end{figure}

To distinguish Rayleigh-like and Love-like phononic modes, we first measure the displacement distributions cross the waveguide (bottom of Figs.$\,$\ref{fig2}(c-d)) with excitation frequency of 102 MHz and 103.5 MHz, respectively. Then, we perform numerical simulation of the phononic modes in the waveguide along $\left[11\bar{2}0\right]$
direction by the three-dimensional finite-element method (COMSOL Multiphysics, V4.3). Figures.$\,$\ref{fig2}(c-d) present simulated displacement distribution of Rayleigh-like and Love-like modes, respectively. Here the x- and z-displacement denote the shear horizontal (in-plane)
and shear vertical (out-of-plane) components, respectively, and all displacement field components are normalized with respect to the maximum of the z-displacement. We found that the Rayleigh-like mode has the displacement mainly in z-direction, and the maximum displacement is located at the center of the waveguide and agrees well with our experiment results in the bottom panel of Fig.$\,$\ref{fig2}(c). In contrast to the case of Rayleigh-like mode, the displacement field profile in the bottom panel of Fig.$\,$\ref{fig2}(d) shows a superposition of $z$-direction displacement field component of the Love-like mode and the Rayleigh-like mode, indicating that both phononic modes are excited simultaneously at the excitation frequency of 103.5 MHz. The
unexpected excitation of Love-like mode, which is piezoelectrically inactive in a bulk GaN crystal \citep{GaN}, is attributed to the mode hybridization in a phononic wire waveguide.

To further verify the mode hybridization in the phononic waveguide, we perform a detailed characterization of the mode profile  along $\left[11\bar{2}0\right]$ direction with excitation frequency fixed at 103.5$\,$MHz. The two-dimensional (2D) displacement distribution ($Z_{0}(x,y)\mathrm{sin}(\Omega t+\phi)$
with fixed $t$ ) of the sample is constructed from the experimental data of recorded amplitude and phase $\left(Z_{0}\,\phi\right)$, as shown in Fig.$\,$\ref{fig3}(a). Fig.$\,$\ref{fig3}(b) shows
the displacement along y-direction approaching the center of the waveguide ($x=240\:\textrm{\ensuremath{\mu}m})$. While the 2D image indicates excellent phonon confinement in the waveguide, the measured field is not uniform along the propagation direction. This can be explained by the mode hybridization effect, the propagating phononic field is composed of a superposition of Rayleigh-like and Love-like modes. As shown in Fig.$\,$\ref{fig3}(b), the experimental data can be fitted by the beating of two waves with different propagation phase velocities. Additionally, the observed asymmetric field distribution at the cross-section of the waveguide in bottom panel of Fig.$\,$\ref{fig2}(d) is also fitted by such a superposition, further confirming the mode
hybridization effect. Shown in Fig.$\,$\ref{fig3}(c) is the calculated out-of-plane displacement of the waveguide, which matches the experimental results. For the guided Love-like mode, although the out-of-plane displacement is not the dominant component of vibration, its non-zero value still permits efficient optical readout of the Love-like modes in our experiment. Here, the simulated displacement is a superposition of $z$-direction motion of the Love-like mode and the Rayleigh-like mode along the waveguide, with the amplitude ratio is around $2.8:1$.

\begin{figure}
\begin{centering}
\includegraphics[width=1\columnwidth]{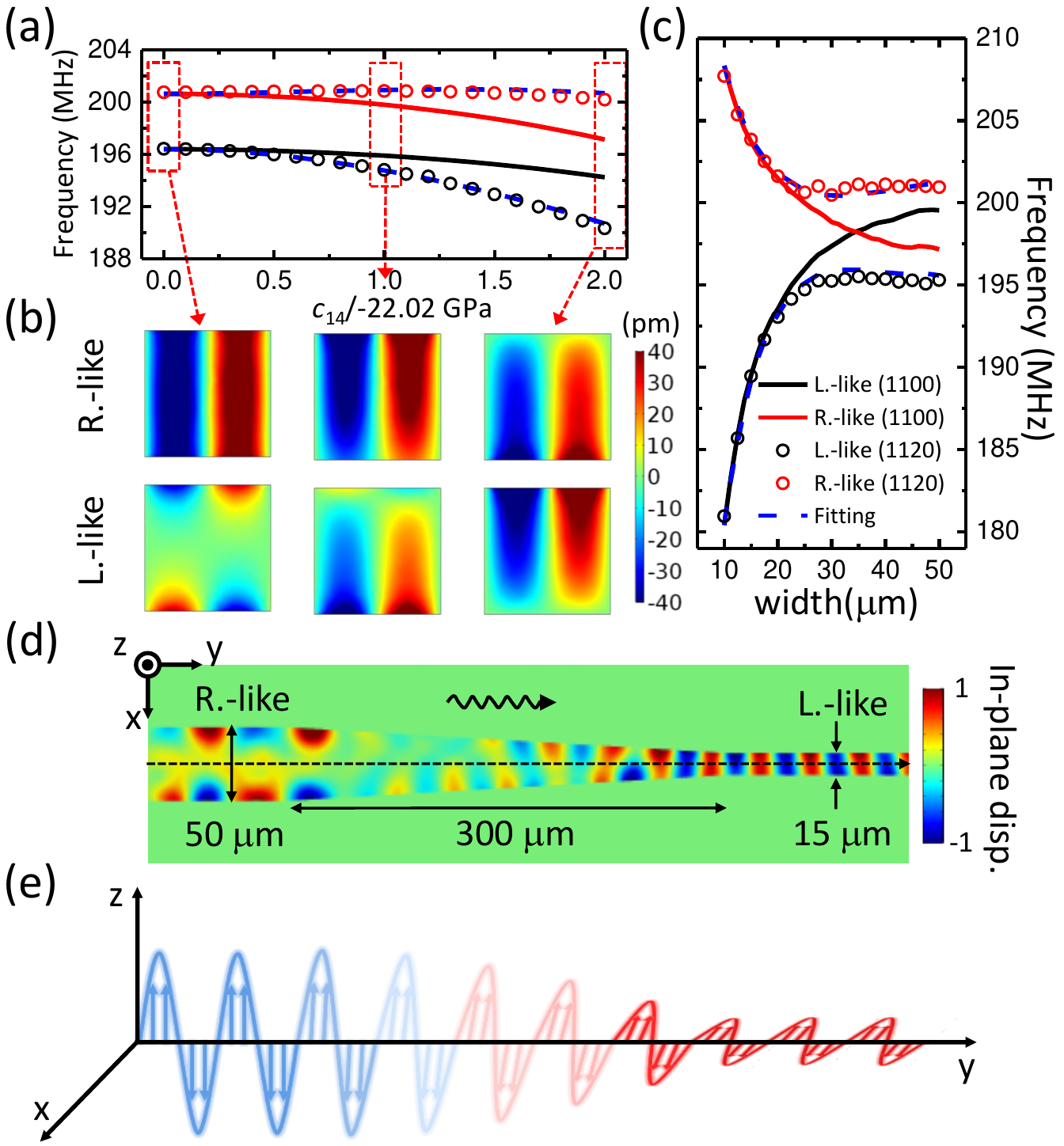}
\par\end{centering}
\caption{(a) Calculated modal frequencies of the Love-like and Rayleigh-like modes supported by a wire waveguide with varying substrate elastic coefficient $c_{14}$. Here, The width of waveguide and the wavelength of phonon are both $25\:\textrm{\ensuremath{\mu}m}$. (b) The simulated
top-view patterns of Love-like and Rayleigh-like modes for $c_{14}=0$, -22.02 GPa, -44.04 GPa, respectively. (c) Calculated frequencies of Love-like and Rayleigh-like modes as a function of the waveguide width. In (a) and (c), the black dotted, red dotted and dash blue lines represent Love-like mode, Rayleigh-like mode and the fittings along the $\left[11\bar{2}0\right]$ direction, respectively. The solid curves are modal dispersions for $\left[1\bar{1}00\right]$-oriented waveguide.  (d) Top view of simulated in-plane displacement of a tapered phononic waveguide. The input Rayleigh-like mode is converted into
Love-like mode after passing through the tapered waveguide due to the anisotropic substrate induced mode coupling. (e) A parallel linecut of (d), shows the schematic of polarization conversion.}

\label{fig4}
\end{figure}

As demonstrated above, the Love-like mode, which is normally piezoelectrically inactive \citep{GaN/Sapphire}, can be excited in our experiment, due to the anisotropic substrate induced mode coupling between Love-like mode and the piezoelectically active Rayleigh-like mode. In the following, we further study the mode coupling in another waveguide of $25\:\textrm{\ensuremath{\mu}m}$-width with an IDT period of $25\:\textrm{\ensuremath{\mu}m}$, which matches the wavelength of phonon and corresponds to a frequency of around 200 MHz. To verify the mode coupling mechanism, we first perform the numerical simulation of the $\left[11\bar{2}0\right]$ (circle dots) and $\left[1\bar{1}00\right]$ (lines) modal frequencies of the Love-like mode and Rayleigh-like mode, as shown in Fig. \ref{fig4}(a). Since the elastic coefficient $c_{14}$ of the sapphire substrate induces the hybridization of the Love-like and Rayleigh-like modes, we vary $c_{14}$ in our simulation
from 0 to 2 times of real material parameter of $-22.02\:\textrm{GPa}$. For the $\left[1\bar{1}00\right]$ direction, there is no coupling between Love-like and Rayleigh-like modes, but their frequencies slightly reduce with $c_{14}$ due to the changed elastic constant.
In contrast, the mode frequencies for the $\left[11\bar{2}0\right]$ direction change significantly, due to the non-zero $c_{14}$ induced coupling. We can simply assume the coupling strength between the two mode $g\propto c_{14}$, then the hybridized mode frequencies can be solved as
\begin{equation}
\Omega_{\pm}=\frac{1}{2}\left(\omega_{L,0}+\omega_{R,0}\pm\sqrt{\left(\omega_{L,0}-\omega_{R,0}\right)^{2}+4g^{2}}\right),
\end{equation}
where $\omega_{L,0}$ and $\omega_{R,0}$ are the Love-like and Rayleigh-like modes frequencies along the $\left[1\bar{1}00\right]$ direction. The coupled mode model tell us that the frequency difference between Love-like and Rayleigh-like modes increases with $c_{14}$, which agree with the numerical results (circle dots in Fig. \ref{fig4}(a)).
In Fig.$\,$\ref{fig4}(b), we show the corresponding eigenmode profiles for three coupling configurations: $c_{14}=0$ (no coupling), $-22.02$ GPa (sapphire) and $-44.04$ GPa (strong hybridization). It can be seen that as $c_{14}$ increases from 0 to a finite value, the Rayleigh-like/Love-like mode evolves from pure modes to hybridized modes. In contrast, for
modes propagating along the $\left[1\bar{1}00\right]$ direction, the mode patterns remain unchanged when $c_{14}$ is varied. The mode coupling can be further investigated by varying the waveguide width. Figure \ref{fig4}(c) shows a clear avoided crossing between different modes when varying the width of the $\left[11\bar{2}0\right]$-oriented waveguide. However, for $\left[1\bar{1}00\right]$ waveguides, the dispersion curves can cross each other without coupling due to the absence of mode coupling. All the numerical results in the $\left[11\bar{2}0\right]$ direction can be fitted well (dash blue lines in Figs. \ref{fig4}(a) and (c)) by using Equation (1) with $g=\xi c_{14}[1-exp(-\sigma/\sigma_{0})]$,
where $\sigma$ is the width of waveguide, with the fitting parameters
$\xi=307\:\textrm{Hz}/(\textrm{Pa}\cdot\textrm{m})$ and $\sigma_{0}=10\:\mu\textrm{m}$.
Here, the exponential dependence on $\sigma$ is due to the fields of Love-like mode in substrate exponentially increase with the $\sigma$, according to the numerical simulation results\citep{Wance}. This mode coupling can be also clearly visualized in the supplementary material videos constructed from the amplitude and phase measurements. Movie (1) shows the measured result at an excitation frequency of 188 MHz. The hybridized Rayleigh and Love modes were excited simultaneously. Note the scanned range is $20\:\mu\textrm{m}$,
which is slightly smaller than the width of waveguide of $25\:\mu\textrm{m}$. Movie (2) shows the simulated field distributions as a superposition of the profiles of
Rayleigh-like and Love-like modes.

The substrate anisotropic-induced mode coupling makes it possible to realize flexible control of the supported phononic modes in a phononic circuit. For example, we can design a tapered waveguide to realize adiabatic Rayleigh-like to Love-like mode conversion. As a proof of concept, a simulation of such a mode converter is conducted in Fig.$\,$\ref{fig4}(d), where the in-plane displacement is shown. It is shown that Rayleigh-like mode is excited by IDT at the input port and adiabatically evolved into the the Love-like mode after passing through the tapered waveguide. A parallel linecut of the tapered waveguide shows the schematic of polarization conversion in Fig.$\,$\ref{fig4}(e).    
Using adiabatic mode conversion, we can efficiently actuate the Love-like mode by the IDT.

In conclusion, the hybridization of Rayleigh-like and Love-like modes in a GaN phononic waveguide grown on anisotropic sapphire substrate is experimentally demonstrated. The mechanism of the mode hybridization is the anisotropic elastic coefficient $c_{14}$ of the substrate induced coupling between in-plane and out-of-plane mechanical motions. Based on this mechanism, a tapered phononic waveguide can be designed to realize mode conversion from Rayleigh-like mode to Love-like mode. The efficient electrical excitation and optical measurement of Love-like mode could be used for study of Love-like mode resonators
with improved quality factors and high-sensitive detection in ambient
\citep{Wei}.

\textbf{Acknowledgment} Zhen Shen and Wei Fu contributed equally to
this work. Z.S. thanks X.B. Xu and C.H. Dong for helpful discussions.
This work is supported by DARPA/MTO's PRIGM:AIMS program through a
grant from SPAWAR (N66001-16-1-4026), an Army Research Office grant (W911NF-14-1-0563), an
Air Force Office of Scientific Research (AFOSR) MURI grant (FA9550-15-1-0029).
H.X.T. acknowledges support from a Packard Fellowship in Sceince and
Engineering. The authors thank Michael Power and Dr. Michael Rooks
for assistance in device fabrication.

\end{document}